\documentclass{llncs}

\usepackage{subfig}
\usepackage{graphicx}
\usepackage{multirow}
\usepackage{subfig}
\usepackage{tabularx}

\begin{document}

\title{The Scalability of Trustless Trust}
\titlerunning{The Scalability of Trustless Trust}
%
\author{Dominik Harz\inst{1} \and Magnus Boman\inst{2,3}}
\authorrunning{Dominik Harz and Magnus Boman} 
%
\tocauthor{Dominik Harz and Magnus Boman}
\institute{IC3RE, Imperial College London, SW7 2RH London, UK,\\
\and
RISE, Box 1263, SE-16429 Kista, Sweden,\\
\and
KTH/ICT/SCS, Electrum 229, SE-16440 Kista, Sweden}

\maketitle              

\begin{abstract}
Permission-less blockchains can realise trustless trust,
albeit at the cost of limiting the complexity of computation tasks.
To explain the implications for scalability, we
have implemented a trust model for smart contracts, described as agents in an open multi-agent system. Agent intentions are not necessarily known and autonomous agents have to be able to make decisions under risk. The ramifications of these general conditions for scalability are analysed for Ethereum and then generalised to other current and future platforms.
\end{abstract}

\section{Introduction}

Turing-complete programming languages allow creating a generic programmable blockchain by means of smart contracts \cite{Vukolic2016}.
A smart contract can be defined as a decentralised application executed on the distributed P2P network that constitutes the blockchain. 
The smart contract captures the formalisation of electronic commerce in code, to execute the terms of a contract. However, a smart contract is, in fact, neither smart nor a contract. In practice, it codes an agreement about what will come to pass, in the form of a production rule. Since there cannot be a breach of contract---which would happen only if one or more parties would not honour the agreement---thanks to how this production rule is coded, a smart contract is not a contract. Since there is no opportunity for learning on the contract's behalf, it is also not smart. 

Smart contracts do code the preferences of their owners, and their negotiating partners as appropriate, with respect to the decision under risk or uncertainty. They react on events, have a specific state, are executed on a distributed ledger, and are able to interact with assets stored on the ledger \cite{Szabo1997}.
Ethereum offers smart contracts through its blockchain. The Ethereum Virtual Machine (EVM) handles the states and computations of the protocol and can theoretically execute code of arbitrary algorithmic complexity \cite{Buterin2013}.
Using Ethereum, developers can implement smart contracts as lines of code in an account that execute automatically when transactions or function calls are sent to that account.
The outcome is final and agreed on by all participants and blockchains can thus enable a system of trust. 

In Ethereum, smart contracts can interact through function calls via their Application Binary Interface (ABI).
Single smart contracts or multiple smart contracts together can act as decentralised autonomous organisations by encoding the rules of interaction for the organisation's inner and outer relationships (e.g., The DAO, MakerDAO).
\textit{Full nodes} store the distributed ledger and validate new blocks in the chain \textit{pro bono}.
Permission-less blockchains limit the complexity of computation tasks and thus, the scalability of these blockchains.
When utilising smart contracts, external services can be required to circumvent these computational limitations to code the preferences of their owners.
The result of computations performed by external parties are not subject to the consensus protocol of the underlying blockchain, and their provided solution or correct execution cannot be formally verified.
Hence, the oft-cited benefit of blockchains allowing for transparency over every transaction and enforced trust through a consensus mechanism cannot be guaranteed with external entities \cite{Kosba2015}.
A trust model for smart contracts in permission-less blockchains is thus missing, a fact that limits their adaptability. Earlier trust models used in related applications, such as those devised for quantitative trading or speculative agent trading (see the patent text \cite{Hoffberg2007} for a good indication of this range), need to be adjusted for the inherent transparency and particular trust implications of blockchain systems.
We propose a model that incorporate all these aspects.

\section{Method}\label{method}
We answer the following research questions:
\begin{enumerate}
\item Which models of trust can be applied to smart contracts to reflect public permission-less blockchains?
\item What can be done to clarify the link between, on the one hand, the preferences and intentions of authors of smart contracts and, on the other hand, the run-time properties of those smart contracts?
\item How can properties of trust models be applied to verify computations in permission-less blockchains?
\end{enumerate}

Question 1 is analysed in two steps. 
First, the applicability of agent-based trust models for smart contracts is evaluated by deducing their strong and weak notions based on agent theory. Second, a trust model suitable for smart contracts in permission-less blockchains is developed, based on a review of existing multi-agent system trust models \cite{Pinyol2013}. 
Question 2 is analysed deductively, based on literature on decision theory and decision analysis, and on limitations of formal representations of preference, and their logical closure, e.g., what can be derived from them.
Question 3 is investigated instrumentally, by developing an algorithm for verifiable computations.
The development of the algorithm followed a deductive method of merging verifiable computation concepts using blockchains \cite{Zyskind2016} \cite{Teutsch2017} with cloud and distributed systems research \cite{Canetti2011} \cite{Canetti2013}.
This revolves around preserving privacy of user data, whereby aspects of the blockchain are used to enforce the algorithm \cite{Zyskind2016}, and on
verifiable computation for Ethereum using computation services inside the blockchain \cite{Teutsch2017}. In the latter, a verification algorithm with dispute resolution and an incentive layer were suggested, and the relevant assumptions critically assessed to develop a new algorithm, since their proposal had two practical issues: 
First, the verification game includes a 'jackpot' to reward solvers and verifiers for their work.
This introduces an incentive to steal the jackpot by solvers and verifiers colluding to receive the jackpot without providing a correct solution.
Second, they propose to implement the computation tasks in C, C++, or Rust code using the Lanai interpreter implemented as a smart contract on Ethereum.
This limits the flexibility of computation services by forcing them to use one of the three programming languages. The objective of the here presented algorithm is to achieve:
\begin{enumerate}
\item execution of arbitrary computations requested from a smart contract in Ethereum, and executed outside the blockchain;
\item verification of the computation result achievable within reasonable time, i.e., $\mathcal{O}(n)$;
\item guarantees that the result of the computation is correct without having to trust the providing service.
\end{enumerate}
Our development was experimental and explorative. 
Different parameters and the agents they pertain to were first considered in a pen and paper exercise, then validated via qualitative assessment as well as quantitative analysis.
The quantitative experiments constitute an evaluation basis for the last two algorithm objectives.

\section{Explicating Smart Contracts}
\label{explicate}
Consensus protocols are used to decide upon the state of the distributed ledger
\cite{Narayanan2016}. This ledger is in permission-less blockchains accessible to anyone participating in the network and through blockchain explorers even to entities outside of the network. This means everyone is able to see for example which public key owns the most Ether. Also, each transaction can be inspected, making it possible for participating parties to monitor the progress of their transaction.
To provide an incentive to the miner and prevent unnecessary changes to the ledger, blockchains introduce fees on executing transactions \cite{Narayanan2016}. 
In Ethereum, the blockchain stores transactions and the code of smart contracts as wellas their state. Hence, the state of a smart contract needs to be updated in the same fashion as executing a transaction including fees, consensus, and mining time.
%

Smart contracts on Ethereum are executed by each node participating in the P2P network
and hence operations are restricted to protect the network \cite{Wood2014}.
To circumvent operational issues (e.g., someone executing a denial of service attack on the network), Ethereum introduces a concept to make users pay for execution of a smart contract functions, and the EVM supports only certain defined operations \cite{Wood2014}, with each operation coming with a certain cost referred to as \textit{gas}.
Before executing a state-changing function or a transaction, the user has to send a certain amount of gas to the function or the transaction.
Only if the provided amount of gas is sufficient for the function or transaction to execute, it will successfully terminate. Otherwise, the transaction or function will terminate prematurely, with results contingent on the handling of the smart contract function.

We now look at two ways of explicating the roles that smart contracts may take on. First, the agent metaphor is employed to
provide an informal understanding in terms of a widely 
accepted and understood terminology. Second, the concept of utility is employed to provide a formal understanding of how the preferences and intentions of smart contract owners may be encoded in the contract itself.

\subsection{Smart Contracts as Agent Systems}
\label{chap2:smart-contracts-agent}
Agents have certain properties separable in weak and strong notions \cite{Wooldridge2009}. Weak notions include \textit{autonomy}, \textit{pro-activeness}, \textit{reactivity}, and \textit{social ability}.
\textit{Autonomy} refers to the smart contract ability to operate without a direct intervention of others and include control over their actions and state. In Ethereum, the state of smart contracts is maintained on the blockchain, while the actions are coded into the smart contract itself. These actions can depend on the state, thus providing a weak form of autonomy.
\textit{Pro-activeness} describes goal-directed behaviour by agents taking initiative. This is somewhat limited in Ethereum, as smart contracts currently act on incoming transactions or calls to their functions. However, if one perceives an agent as a collection of multiple different parts, smart contracts might well be extended by external programs triggering such initiatives. Thereby, the limitations set by Ethereum can be circumvented and an agent with pro-active notions can be created. The result is in effect a multi-agent system and can be analyzed as such.
\textit{Reactivity} is based on perception of an agent's environment and a timely response to those changes. By design, smart contracts only have access to the state of the blockchain they are operating in. Reactivity for state changes in Ethereum is reached via event, transaction, or function implementation. To react to environment changes outside of the blockchain (e.g. executing a function based on changes in stock market prices) requires importing this information to the blockchain via e.g. Oracles \cite{Buterin2016}.
\textit{Social ability} enables the potential interaction with other agents or humans through a communication language. In Ethereum, users and contracts are identifiable by their public key \cite{Wood2014} and interaction is possible through transactions or function calls on smart contracts. 

Strong notions include properties such as \textit{beliefs and intentions}, \textit{veracity}, \textit{benevolence}, \textit{rationality}, and \textit{mobility}. As mentioned in the introduction above, pro-activeness is somewhat limited in Ethereum smart contracts, and so these properties are present only to a limited extent.
The two properties \textit{veracity}, which refers to not knowingly communicating false information, and \textit{rationality}, describing the alignment of the agent's actions to its preferences, 
both pertain to the
incentives an author of a smart contract might have to develop an agent which is rational but not truthful, in order to maximise profits. This can be deliberate so that the agent correctly encodes the true preferences of the smart contract owner, or non-deliberate, in which case the owner preferences might be inadequately coded.
To deal with the uncertainty of agent intentions, three approaches have emerged. 
First, security approaches utilise cryptographic measures to guarantee basic properties such as authenticity, integrity, identities, and privacy \cite{Pinyol2013}. Within blockchains, this is mainly achieved through cryptographic measures, which do not provide trust in the content of the messages.
Second, institutional approaches enforce behaviour through a centralised authority. This entity controls agents' actions and can penalise undesired behaviour. Governance functions enforcing behaviour not defined in the core protocol do not exist.
Third, social approaches utilise reputation and trust mechanisms to e.g. select partners, punish undesired behaviour, or evaluate different strategies. 
In blockchains, 
there is no system of trust implemented in the core protocol, which would rate behaviour according to certain standards.
These three approaches are complementary and can be used to create a system of trust \cite{Pinyol2013}. Trust research and current implementations are primarily focused on the first two approaches. This allows creating agents on a platform that enforces these defined trust measurements \cite{Balakrishnan2013} \cite{Sabater2005} \cite{Ramchurn2004} \cite{Mui2002}.

\subsection{Utility and Risk}
\label{utility}

Some researchers believe that all game-theoretical aspects of making decisions can be pinned down by logical axiomatizations: it is only a matter of finding the right axioms. Game-theoretical studies often concentrate on two-person games, one reason being that many conflicts involve only two protagonists. In any game, the players may or may not be allowed to cooperate to mutual advantage. If cooperation is allowed, the generalized theory of $n$-person games can sometimes be reduced to the one for two-person games, since any group of cooperating players may be seen as opposing the coalition of the other players. In the case of smart contracts, this would allow for an owner of multiple contracts (in effect, a multi-agent system) to maximize the utility of interplaying contracts by employing game theory, at least on paper.
For a given set of smart contracts, the problem is how to determine a rule that specifies what actions would have been optimal for the smart contract owner. Actions could here pertain to details of a particular contract, or to the order of their execution, for instance. Comparing different rules measures the risk involved in consistently applying a particular rule, e.g., a chain of smart contract employment. Formally, we wish to determine a decision function that minimizes this risk. 
The simpler case of handling risk is in decisions under certainty. This means that the owner of one or more smart contracts can predict the consequences of employing them. This represents the ideal case in which all smart contracts execute as intended. Thus, the owner simply chooses the alternative whose one and only possible consequence has a value not less than the value of any other alternative. This seems simple enough, but it is necessary to investigate a bit further what the value of a consequence denotes. The preferences of the owner should be compatible with the following axioms ($A$ is not preferred to $B$ is henceforth denoted by $A \leq B$).

$\leq$ is a weak ordering on the set of preferences P:\\
A1. (i)	Transitivity: If $A \leq B$ and $B \leq C$, then $A \leq C$, for all A, B, and C in P.\\
A1. (ii) Comparability: $A \leq B$ or $B \leq A$, for all A and B in P.\\
From this, we may derive the relation of indifference and strict preference, and we state the consistency criteria for these:\\
A2.	(i)	$A = B$ is equivalent to $A \leq B$ and $B \leq A$, for all A and B in P.\\
A2.	(ii) $A \leq B$ is equivalent to $A \leq B$ and not $B \leq A$, for all A and B in P.\\
However, A1 implies that the owner has to admit to all consequences being comparable. This is typically not the case in smart contracts, and it becomes necessary to replace Comparability with Reflexivity, yielding a partial ordering instead:\\ 
A1. (iii) Reflexivity: $A \leq A$, for all A in P.

There is much to be gained by representing the preference ordering as a real-valued order-preserving function. If we cannot find such a function there is not much sense in speaking of the numerical value of a sequence of employed smart contracts, and we might as well throw a coin for deciding. Assuming axioms A1 and A2 hold, we must find a function $f(X)$ with the property $f(A) \leq f(B)$ iff $A \leq B$, which we can always do fairly easily for decisions under certainty \cite{French1986}, but we now turn to decisions under risk, which is the class of decisions that normally pertain to owners of smart contracts. In the Bayesian case, with subjective probabilities, we can think of a smart contract employment S as consisting of a matrix of probabilities $p_1 ,..., p_n$ and their corresponding consequences $c_1 ,..., c_n$. Then the real-valued function $f(X)$ we seek lets us compute the value of S as $\Sigma p_i f(c_i)$.
This fixes one possible definition of an agent as rational, by making it maximize its own utility (in accordance with its preferences, i.e. with the preferences it codes). Formally, an agent accepts the utility principle iff it assigns the value $\Sigma p_i v_i$ to S, given that it has assigned the value $v_i$ to $c_i$. Any ordering $\Omega$ of the alternatives is compatible to the principle of maximizing the expected utility iff $a \Omega b$ implies that the expected value of $a$ is higher than the expected value of $b$. In other words, we are now free to start experimenting with various axiom systems for governing the owners, or at least recommending them actions based on the smart contracts they have at hand. While game-theoretic axiom systems have been favoured among agent researchers, a wide variety of axiomatizations are surveyed in the more formal literature \cite{Fishburn1989} \cite{Malmnas1994}.

\section{A Trust Model for Smart Contracts}
From the 25 models covered in \cite{Pinyol2013}, five consider global visibility and nine consider cheaters. The overlap of those models leaves one model focusing on reputation of actors in electronic markets \cite{Rasmusson1996}. The core idea is to use incentives to encourage truthful behaviour of agents in the system by social control. Social control implies that actors in the network are responsible for enforcing secure interactions instead of using an external or global authority. 

Assuming a rational agent, there is a possible motivation to break protocol if this maximizes utility. Speculation-free protocols have been recommended for some agent applications, but the Ethereum smart contract environment is much too complex to allow for such control features, which require equilibrium markets \cite{Ygge1997}.
To provide a certain level of trust, new agents have to deposit a certain cryptocurrency value for participation, and this deposit is returned when an agent decides to stop participating. However, dishonest or corrupt agents can be penalised by either destroying their deposit or distributing it to honest agents. This is in line with norm-regulation of agent systems \cite{Boman1999} and does not make any other strong requirements on models. Norm-regulation has been formalized for multi-agent systems, e.g., in the form of algebra \cite{Odelstad2004}.

Gossiping can be used to communicate experiences with other agents in a P2P fashion and thereby establish trust or reputation. In the protocol of Bitcoin or Ethereum gossiping is the basis for propagating new transactions and subsequently validating blocks \cite{Decker2015}.
A similar approach can be taken for smart contracts, whereby agents could exchange knowledge or experiences of other agents \cite{Carboni2015}. Reputation of an agent is based on its interaction with other agents, whereby agents mutually need to sign a transaction if they are satisfied with the interaction. Over time, an agent collects these signed transactions to build up its reputation. However, this model is prone to colluding agents boosting their reputation \cite{Can2013}. 
Trust can also be implemented by relying on independent review agents \cite{Huynh2006} \cite{Jakubowski2010} \cite{Cerutti2013a}. 
However, both gossiping and review agents are subject to detection rate issues.  

\section{Applying Trust Measures to Verifiable Computation}
\begin{figure}[h]
\centering
\includegraphics[width=0.7\textwidth]{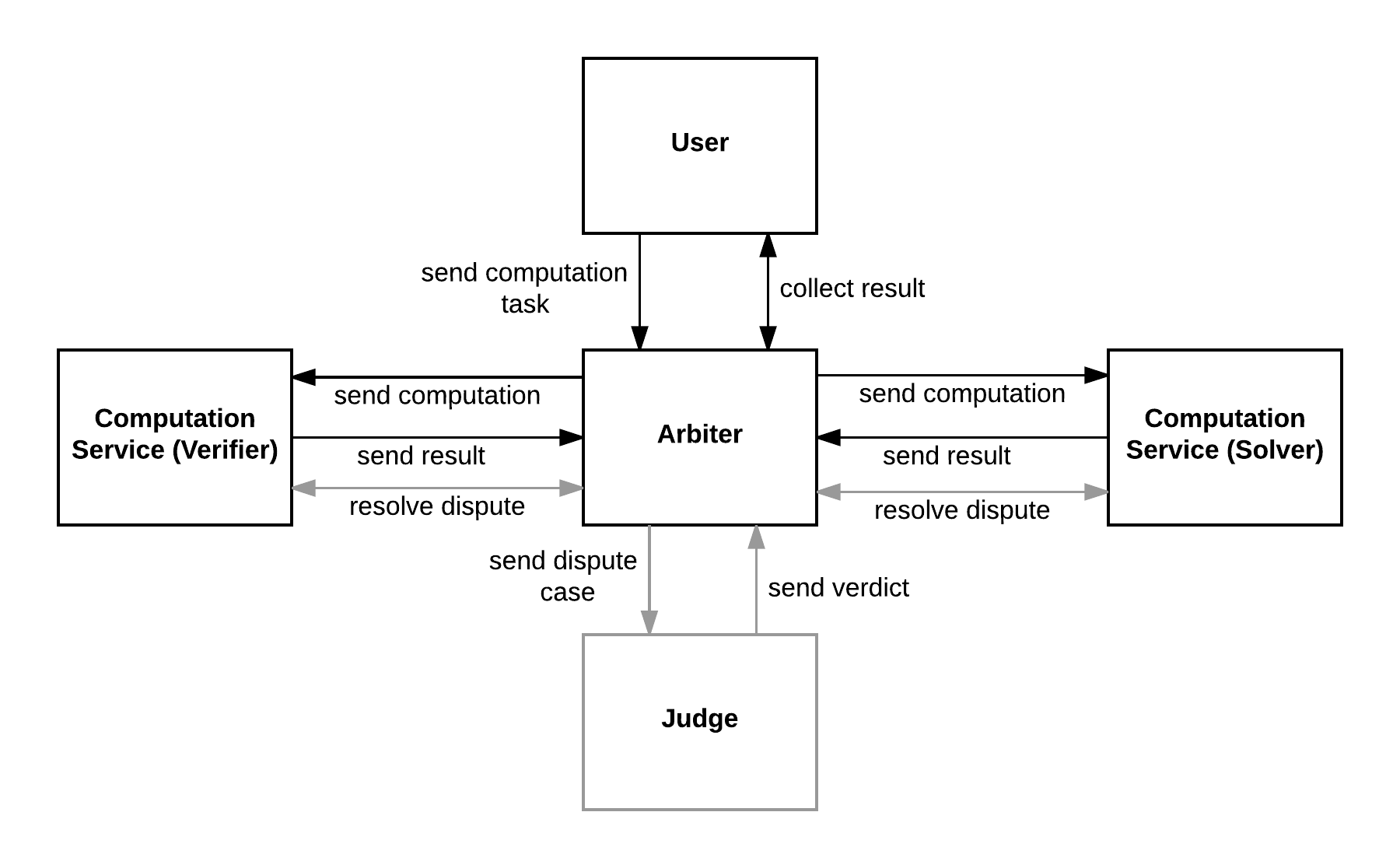}
\caption{Overview of actors in the verification algorithm.}
\label{fig:actors}
\end{figure}

Due to the restrictions set by the EVM (i.e. gas cost of operations), implementing functions in Ethereum with a complexity greater than $\mathcal{O}(n)$ is not feasible. 
To circumvent these limitations, computations can be executed outside of Ethereum and results stored on the blockchain.
We present an algorithm to achieve verifiable computations outside of Ethereum through measures presented in the trust model. 
Agents' rational behaviour can be aligned to the overall objective of the algorithm.
The actors involved in the verifying computation algorithm are presented in Fig.\ref{fig:actors}.
\textit{Users} request solving a specific computation problem.
They provide an incentive for solving and verifying the problem.
\textit{Computation services} provide computation power in exchange for receiving a compensation.
For participation, they are providing a deposit.
One of the computation services acts as a \textit{solver} and at least one other computation service acts as a \textit{verifier}.
\textit{Judges} decide whether basic mathematical operations are correct or not.
They are neutral parties and are not receiving any incentives.
An \textit{arbiter} enforces the verifiable computation algorithm when users request a new computation.

Users are assumed as agents with the objective to receive a correct computation. They are required to send a fee to reward solvers and verifiers for executing the computation. This fee depends on the complexity of the computation to be performed, the complexity of the input data, and the number of verifiers.
Computation services are assumed to optimise their incentive. They might purposely communicate false information to maximise their incentive. Further, enough computation services are available (i.e. a minimum of 2) to execute the computation with at least one verifier. The probability of detecting a false computation depends on the number of verifiers in the algorithm.
The arbiter and judge are trusted by participating parties, respectively enforcing the algorithm and reaching a verdict. This is a strong assumption in a trustless system and needs to be justified. To limit their incentive for undesired behaviour (i.e. cheating) in the algorithm, these two agents are not rewarded for taking part in the computations. Thus, their work is \textit{pro bono} and only the operational cost in gas are covered.

Alternatively and not further covered in this paper, other approaches limit or eliminate trust in arbiter and judge. First, following the trust is risk approach \cite{Litos2017}, a network of trusted entities with a fixed amount of deposited value could be created to find arbiters and judges trusted commonly between computation services and users. Second, a user might create their own arbiter and judge, while storing the fee in an escrow contract between user and computation services. The computation services store an encrypted hash of the result in the escrow contract. Upon completion of the protocol, the user issues the payment and receives the result in full. Third, the protocol could be executed with different test cases while results would be publicly stored on the blockchain. Thus, a user and computation service could verify correct execution of the protocol, if arbiter and judge remain unchanged.

\subsection{Algorithm}
The algorithm is initiated when a user requests a computation by sending the input data, the operation to be performed, and the desired number of verifiers to the arbiter. 
One computation service is randomly determined as a solver, and the other(s) are randomly assigned as verifiers by the arbiter.
The user instructs the arbiter to forward the input data and operation to the computation services smart contracts, triggering the off-chain computation by sending a request through an oracle.
This requires sending a fee for the computation as well as providing the fee for using the oracle.
Verifiers and the solver report their result back to the arbiter. 
If all results are reported back, then the user can trigger the arbiter to compare the available results. If the solver and all participating verifiers agree on one solution, the algorithm is finished and the user can collect the result. 
However, if at least one verifier disagrees with the solver the user can initiate a dispute resolution algorithm. The dispute resolution is inspired by a technique introduced in \cite{Canetti2013}, \cite{Canetti2011}, and \cite{Teutsch2017}: to split up the operation into simple parts with intermediary results until the computation is simple enough for the judge to solve it. Overall and intermediary results are stored in a Merkle tree for the solver, and each verifier challenging the solver. The comparison is achieved through a binary search on the trees. The root of the tree encodes the overall result, while the leaves in the lowest layer encode the input data. Leaves in between represent intermediary results.

\subsection{Interactions}\label{algorithm:scenarios}
Under the assumption that arbiter, judge, and user behave rational and follow the algorithm, computation services have a combination of four different behaviours with respect to their role as solver $S$ or verifier $V$.
The behaviours are summarised in Table \ref{algorithm:incentives} with either verifiers accepting the solution (i.e. $V_A$) or challenging the solution (i.e. $V_C$). $S$ profits the most if it provides a correct solution, which is challenged by $V$, while $V$ profits the most when $S$ provides a false solution and $V$ is able to challenge it. The problematic case is that the incentives for accepting a false or correct solution are the same. To prevent this from happening we will consider the behaviour of $V$ and $S$ in detail.

\begin{table}[]
\centering
\caption{Possible behaviours of computation services as solver $S$ and verifier $V$, whereby all verifiers behave the same.}
\label{algorithm:incentives}
\begin{tabular}{llll}
 &  & \multicolumn{2}{c}{\textbf{$S$}} \\
 & \multicolumn{1}{l|}{} & \multicolumn{1}{c|}{correct solution} & \multicolumn{1}{c}{false solution} \\ \cline{2-4}
\multicolumn{1}{c}{\multirow{2}{*}{\textbf{$V$}}} & \multicolumn{1}{c|}{challenge} & \multicolumn{1}{l|}{\begin{tabular}[c]{@{}l@{}}$S$ receives $S$ fee share\\ $S$ receives $V_C$ fee share\\ $V_C$ receives nothing\end{tabular}} & \begin{tabular}[c]{@{}l@{}}$S$ receives nothing\\ $V_C$ receives $V_C$ fee share\\ $V_C$ receives $S$ fee share\end{tabular} \\ \cline{2-4}
\multicolumn{1}{c}{} & \multicolumn{1}{c|}{accept} & \multicolumn{1}{l|}{\begin{tabular}[c]{@{}l@{}}$S$ receives $S$ fee share\\ $V_A$ receives $V_A$ fee share\end{tabular}} & \begin{tabular}[c]{@{}l@{}}$S$ receives $S$ fee share\\ $V_A$ receives$V_A$ fee share\end{tabular}
\end{tabular}
\end{table}

\textbf{Case 1:} $S$ provides a correct solution and no $V$ challenges the solution. Agents behave as intended by the algorithm. As no $V$ challenges the solution, the judge is not triggered and the fee is equally split between $S$ and the involved $V$.

\textbf{Case 2:} $S$ provides a correct solution and at least one $V$ challenges the solution. This is an undesired behaviour since the solution provided is actually correct. This triggers the dispute resolution with a verdict by the judge determining $S$ as correct. In this case $S$ profits from the extra work due to the additional dispute steps by receiving the fee share of $V_C$. $V_A$ receive their part of the fee since their amount of work remained the same.

\textbf{Case 3:} $S$ provides a false solution and no $V$ challenges the solution. $S$ and all $V$ would receive their share of the fee. This is an undesired behaviour in the algorithm as it would flag a false result as correct. To prevent this from happening two measures are used. First, computation services do not know their role in advance as they are randomly assigned by the arbiter.
If several services collude to provide false solutions, all of them would need to work together to provide the ``same wrong'' result. However, if just one $V_C$ exists, it profits by gaining the fee shares of itself, $S$, and all $V_A$. Thus, second, the user is able to determine the number of $V$ for each computation. 
The probability of having at least one $V_C$ depends on the prior probability $p$ of $V$ providing correct or false solutions and the number $n$ of $V$ in the computation.

\textbf{Case 4:} $S$ provides a false solution and at least one $V_C$ challenges the solution. Hereby, $S$ and $V_A$ are not receiving their share of the fee, which goes to all $V_C$. This is based on the verdict by the judge. However, this is also an undesired case since the user does not receive a solution to his computation.

Considering the four scenarios, rational $S$ is trying to receive its share of the incentive and get a chance to receive fees of any $V$ challenging a correct solution. The strategy for $S$ considering $V$ is to provide a correct solution to the problem. $V$ profits the most form challenging a false solution. A rational $V$ provides the correct solution to a computation to receive its fee share or to have the chance of becoming a challenger to a false solution.
Arguably, $S$ and $V$ could try to deliver a false solution to save up on computation cost or trick the user. In this case, the probability of discovering the false solution relies on the number of $V$s and the prior probability of cheating $V$s. If a $V$ delivers a false solution, it must be the same solution as $S$' to not trigger the dispute resolution. 
Moreover, by destroying the services' deposits and excluding them from the algorithm after detected cheating, the prior probability of having such a service can be reduced.

\subsection{Implementation and experiments}\label{algorithm:implementation}
The algorithm was implemented using Solidity smart contracts and AWS Lambda external computation services. The quantitative analysis is conducted by executing experiments with one exemplary type of computation. The computation is a multiplication of two integers to simplify the verification steps in the algorithm. The results depend on external and internal parameters of the algorithm. Externally, the prior probability of computation services providing false solutions is considered. Internally, the number of verifiers the user requests for each computation are examined. Experiments are executed for each different configuration of parameters to determine gas consumption and outcome of the computation. Assuming a potentially large number of computation services ($> 10,000$), this gives a confidence level of 95\% and a maximum confidence interval of $3.1$ for the three different prior probabilities. Before each iteration of the experiment, the environment is initialised with a new set of smart contracts. Experiments are executed within \textit{TestRPC} \cite{Ethereum2017b}.

\begin{figure}[h]
\centering
\subfloat[30\% of computation services providing incorrect solutions.]{\includegraphics[width=0.47\textwidth]{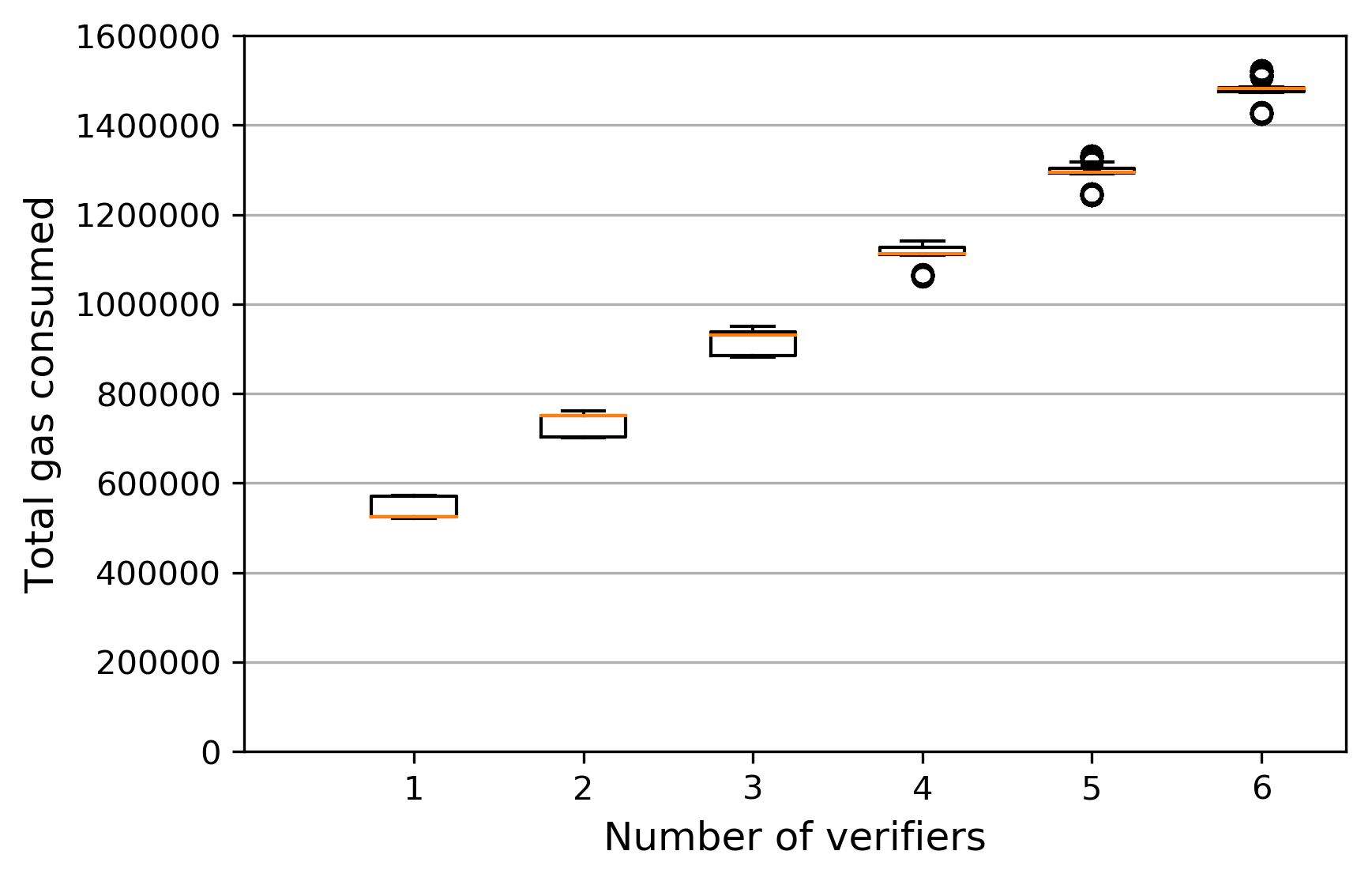}}
\qquad
\subfloat[50\% of computation services providing incorrect solutions.]{\includegraphics[width=0.47\textwidth]{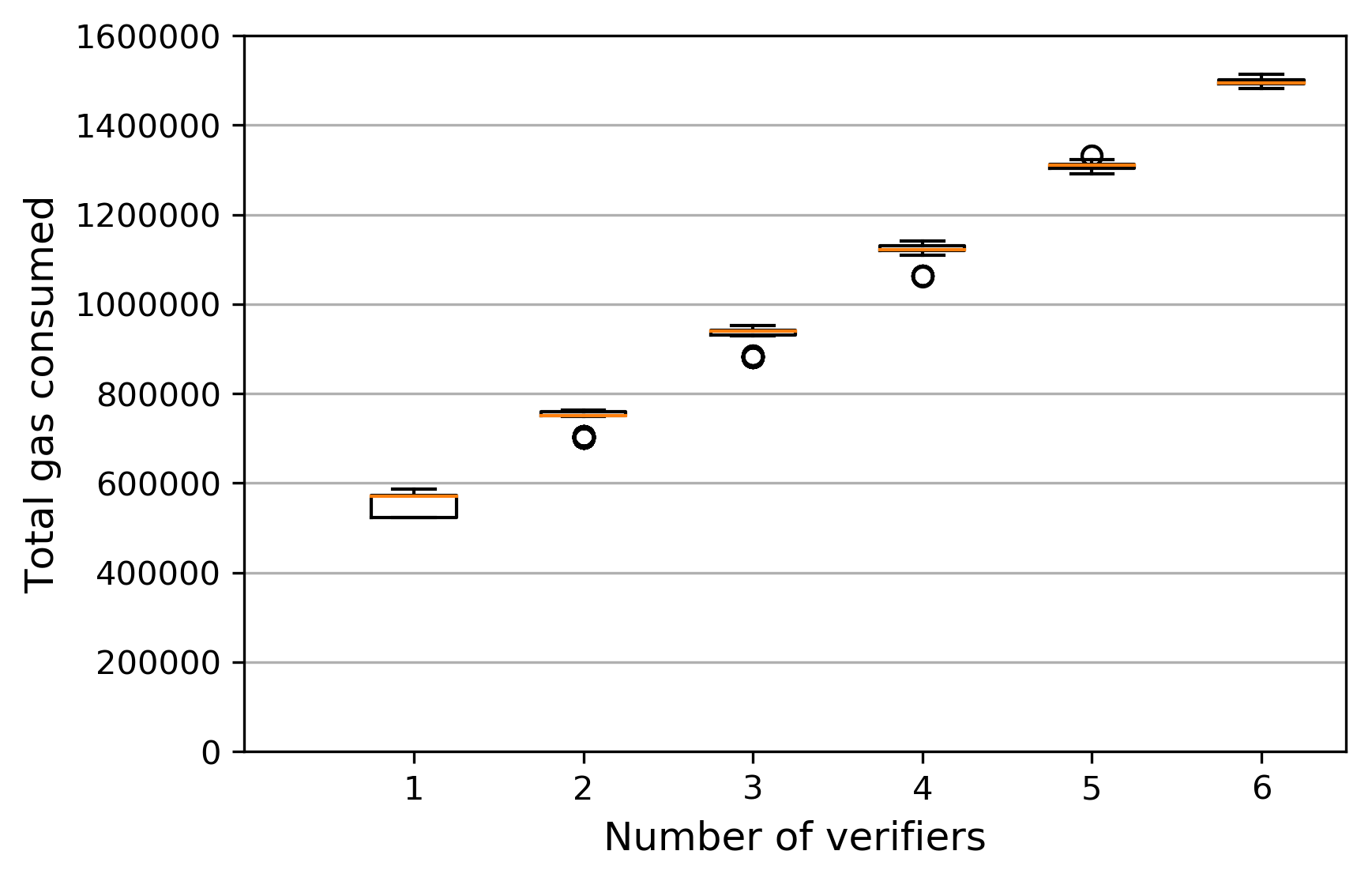}}
\qquad
\subfloat[70\% of computation services providing incorrect solutions.]{\includegraphics[width=0.47\textwidth]{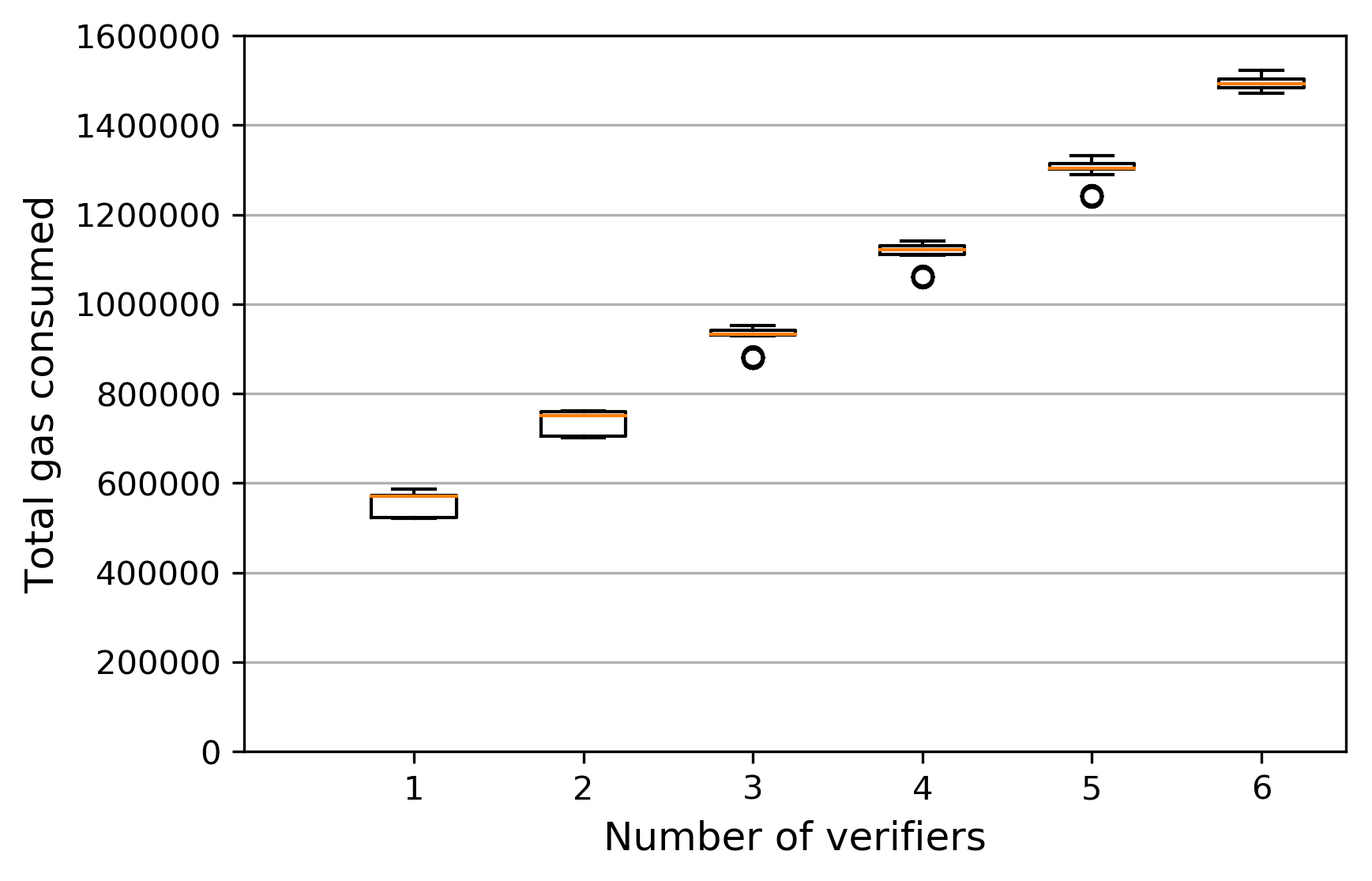}}
\caption{Total amount of gas used by algorithm with different number of verifiers and percentage of computation services providing incorrect solutions. Each combination of specific number of verifier(s) and percentage of computation services with incorrect solutions with $N=1000$.}
\label{fig:gas}
\end{figure}

Reporting the amount of gas used equals the time and space complexity of the algorithm, as gas consumption is determined by the type and number of operations in the EVM. It further excludes the time used for sending transactions or calls. Independent of the prior probability of false solutions, the $\mu$ gas consumption increases linearly as presented in Fig.\ref{fig:gas}. Further, $\sigma$ decreases with an increasing number of verifiers. At a low number of verifiers, the dispute resolution is less likely triggered, leading to a higher $\sigma$ in gas consumption. With an increasing number of verifiers, the probability of triggering the dispute resolution increases. As the dispute resolution is almost always triggered, $\sigma$ is reduced.

The algorithm is tested for three different cases of verification: First, the algorithm can accept a correct solution. Second, each verifier agrees with the solver although the solution is not correct. The dispute resolution is not triggered and the user receives a false solution marked as correct. Third, at least one verifier disagrees with the solver providing a false solution and the judge rules that the solver's solution is false. 
For the second case, invoking the dispute resolution depends on the prior probability of computation services providing false solutions described by $P(V_C) = 1 - p^n$. The experiments as shown in Table \ref{discuss:probabilities} indicate that the expected and actual value are similar for $p = 0.5$. However, for $p=0.3$ and $p=0.7$ the actual values are below the expected ones. Since the experiment is executed with a confidence level of 95\% and interval of $3.1$, those changes are accounted towards sampling size not being a perfect representative of the actual distribution. Also, the random assignment of false and correct computation services could be a cause for having a higher detection rate.

\begin{table}[]
\centering
\caption{Comparison of expected and actual probabilities of accepting a false solution in the algorithm.}
\label{discuss:probabilities}
\begin{tabular}{c|c|c|c}
\hline
\textbf{Prior $p$} & \textbf{Verifiers $n$}  & \textbf{Expected false [\%]}  & \textbf{Actual false [\%]} \\ \hline
0.3 & 1 & 9.0 & 2.7 \\
0.3 & 2 & 2.7 & 0.0 \\
0.3 & 3 & 0.81 & 0.0 \\
0.3 & 4 & 0.243 & 0.0 \\
0.3 & 5 & 0.0729 & 0.0 \\
0.3 & 6 & 0.02187 & 0.0 \\  \hline
0.5 & 1 & 25.0 & 28.6 \\
0.5 & 2 & 12.5 & 12.2 \\
0.5 & 3 & 6.25 & 4.6 \\
0.5 & 4 & 3.125 & 1.2 \\
0.5 & 5 & 1.5625 & 0.0 \\
0.5 & 6 & 0.78125 & 0.0 \\ \hline
0.7 & 1 & 49.0 & 41.2 \\
0.7 & 2 & 34.3 & 24.4 \\
0.7 & 3 & 24.01 & 12.1 \\
0.7 & 4 & 16.807 & 4.9 \\
0.7 & 5 & 11.7649 & 2.9 \\
0.7 & 6 & 8.23543 & 0.0 \\      \hline
\end{tabular}
\end{table}

\section{Discussion}
Within the presented trust model, 
deposits are simple to implement in permission-less blockchains that already have a cryptocurrency. However, the deposit value can be volatile. This poses two risks: Either the escrow or independent entity maintaining the deposit may be motivated to steal the deposits, or the deposit value might be so little that its trust-building attribute vanishes. To prevent this, the deposit value could be bound to a fiat currency or a stable asset. The deposit can also be dynamically adjusted and deposits only kept a short time or one iteration of interactions.
Gossiping could be used as a basis to communicate experiences with other agents. In permission-less blockchains, the agents can use a common protocol to exchange this information and use a rating approach \cite{Zhou2008}. Yet, gossiping can be misused by agents to boost their own reputations by executing Sybil attacks.
Review agents can be used that reach a verdict on a specific issue or problem. Their implementation is simple and potential scenarios to manipulate agents' reputations are prevented. However, the judge or review agent needs to be trusted by other agents.
The algorithm is based on its actors and their interaction. The idea of arbiter, judge, user, and computation services is strongly influenced by \cite{Teutsch2017} and \cite{Zyskind2016}. The main differences are in the idea of using a jackpot to reward verifiers as well as the implementation either entirely on Ethereum or using external computation services. Moreover, the algorithm defers from \cite{Zyskind2016} as its goal is to deliver verifiable computations for entities (i.e. users or smart contracts) on the blockchain, while \cite{Zyskind2016} primarily delivers privacy-preserving computations, where blockchain enables the algorithm.

The algorithm cannot guarantee to detect false solutions. It is based on the assumption that solvers and verifiers behave as desired (i.e. delivering correct solutions), as their strategy is aligned with the incentives provided by the algorithm. This assumption is based on game-theoretic properties. The algorithm leaves no dominant strategy considering the interactions in Table \ref{algorithm:incentives}. $S$ can choose either to provide a correct or false solution and $V$ can challenge or accept. Only when considering both agents, a Nash equilibrium exists. If there is a (high) probability that a $V_C$ exists, the only valid strategy for $S$ is to provide a correct solution. Consequently, $V$ in turn has to provide a correct solution, which accepts correct $S$ and challenges false $S$.
In the algorithm, both $S$ and $V$ providing correct solutions gives a Pareto efficient result. If they change their strategy under the assumption that no $V_C$ exists, their utility remains the same. However, a $V$ has an incentive to challenge a false solution, which would increase his utility and reduce the utility of the others. Social welfare considers the sum of all agent's utilities depending on their strategy which can be disregarded in permission-less blockchains since overall the agent wants to optimise his utility independent of the overall utility. Specifically, the overall utility is potentially unknown to an individual agent, since he is unable to determine with certainty the utility of other agents.

\section{Conclusion}
On permission-less blockchains like Ethereum, rational agents through smart contracts code the preferences of their owners. This could motivate maximizing their utility by dishonest behaviour, and hence, further social control mechanisms are required. We have presented a trust model for smart contracts in permission-less blockchains that incorporate state-of-the-art research into deposits, reputation, and review agents for social control.
Trust can be extended to entities outside of permission-less blockchains through applying the trust measures presented in our model. An example application is an algorithm implementing verifiable computation. The model includes users requesting computational tasks, computational services providing solutions and acting either as solver or verifier, arbiters enforcing the algorithm, and judges resolving disputes. Due to the incentive structure and the potential penalty cause by cheating, providing correct solutions to the computation task is a Nash equilibrium. Under the assumption that arbiter and judge are trusted, the algorithm detects false solutions provided based on a probability distribution. The algorithm is realised as Solidity smart contracts and AWS Lambda functions, implementing verification of multiplying two integers. Experiments show that with six verifiers the algorithm detects cheaters with prior probabilities of 30\%, 50\%, and 70\% dishonest computation services. Experiments show that the algorithm performs overall with a linear time and space complexity depending on the number of verifiers.

As future work, we leave eliminating trust requirements regarding arbiter and judge by a fully decentralised algorithm.

\subsubsection{Acknowledgement}
The authors thank Babak Sadighi and Erik Rissanen for comments and discussions, Daniel Gillblad for important support for Magnus Boman's part of the project. Also, the authors thank Outlier Ventures Ltd. for partly funding Dominik Harz' share of the project.

%
%
\bibliography{presubmit}

\begin{thebibliography}{10}
\providecommand{\url}[1]{\texttt{#1}}
\providecommand{\urlprefix}{URL }

\bibitem{Balakrishnan2013}
Balakrishnan, V., Majd, E.: {A Comparative Analysis of Trust Models for
  Multi-Agent Systems}. Lecture Notes on Software Engineering  1(2),  183--185
  (2013)

\bibitem{Boman1999}
Boman, M.: {Norms in artificial decision making}. Artificial Intelligence and
  Law  7(1),  17--35 (1999)

\bibitem{Buterin2013}
Buterin, V.: {A Next-Generation Smart Contract and Decentralized Application
  Platform} (2013), \url{https://github.com/ethereum/wiki/wiki/White-Paper}

\bibitem{Buterin2016}
Buterin, V.: {Chain Interoperability}. Tech. Rep.~1, R3CEV (2016)

\bibitem{Can2013}
Can, A.B., Bhargava, B.: {SORT: A Self-ORganizing Trust Model for Peer-to-Peer
  Systems}. IEEE Transactions on Dependable and Secure Computing  10(1),
  14--27 (2013)

\bibitem{Canetti2011}
Canetti, R., Riva, B., Rothblum, G.N.: {Practical delegation of computation
  using multiple servers}. In: Proceedings of the 18th ACM conference on
  Computer and communications security - CCS '11. p. 445. ACM Press, New York,
  New York, USA (2011)

\bibitem{Canetti2013}
Canetti, R., Riva, B., Rothblum, G.N.: {Refereed delegation of computation}.
  Information and Computation  226,  16--36 (2013)

\bibitem{Carboni2015}
Carboni, D.: {Feedback based Reputation on top of the Bitcoin Blockchain}
  (2015)

\bibitem{Cerutti2013a}
Cerutti, F., Toniolo, A., Oren, N., Norman, T.J.: {Context-dependent Trust
  Decisions with Subjective Logic} (2013)

\bibitem{Decker2015}
Decker, C., Wattenhofer, R.: {A Fast and Scalable Payment Network with Bitcoin
  Duplex Micropayment Channels}. In: Pelc, A., Schwarzmann, A.A. (eds.) Lecture
  Notes in Computer Science (including subseries Lecture Notes in Artificial
  Intelligence and Lecture Notes in Bioinformatics), vol. 9212, pp. 3--18.
  Springer International Publishing, Cham (2015)

\bibitem{Ethereum2017b}
Ethereum: {Ethereum TestRPC} (2017),
  \url{https://github.com/ethereumjs/testrpc}

\bibitem{Fishburn1989}
Fishburn, P.: {Foundations of Decision Analysis: Along the way}. Management
  Science  35,  387--405 (1989)

\bibitem{French1986}
French, S. (ed.): Decision Theory: An Introduction to the Mathematics of
  Rationality. Halsted Press, New York, NY, USA (1986)

\bibitem{Hoffberg2007}
Hoffberg, S.: Multifactorial optimization system and method (Apr~19 2007),
  \url{https://www.google.com/patents/US20070087756}, uS Patent App. 11/467,931

\bibitem{Huynh2006}
Huynh, T.D., Jennings, N.R., Shadbolt, N.R.: {An integrated trust and
  reputation model for open multi-agent systems}. Autonomous Agents and
  Multi-Agent Systems  13(2),  119--154 (2006)

\bibitem{Jakubowski2010}
Jakubowski, M., Venkatesan, R., Yacobi, Y.: {Quantifying Trust} (2010)

\bibitem{Kosba2015}
Kosba, A., Miller, A., Shi, E., Wen, Z., Papamanthou, C.: {Hawk: The Blockchain
  Model of Cryptography and Privacy-Preserving Smart Contracts}. In: 2016 IEEE
  Symposium on Security and Privacy (SP). vol. 2015, pp. 839--858. IEEE (2016)

\bibitem{Litos2017}
Litos, O.S.T., Zindros, D.: {Trust Is Risk: A Decentralized Financial Trust
  Platform.} IACR Cryptology ePrint Archive  2017,  156 (2017)

\bibitem{Malmnas1994}
Malmn\"as, P.E.: {Axiomatic Justifications of the Utility Principle}. Synthese
  99(2) (1994)

\bibitem{Mui2002}
Mui, L., Mohtashemi, M., Halberstadt, A.: {A computational model of trust and
  reputation}. In: HICSS. Proceedings of the 35th Annual Hawaii International
  Conference on System Sciences. vol.~5, pp. 2431--2439. IEEE (2002)

\bibitem{Narayanan2016}
Narayanan, A., Bonneau, J., Felten, E., Miller, A., Goldfeder, S.: {Bitcoin and
  Cryptocurrency Technologies - Draft}. Princeton University Press, Princeton,
  NJ, USA (2016)

\bibitem{Odelstad2004}
Odelstad, J., Boman, M.: Algebras for agent norm-regulation. Annals of
  Mathematics and Artificial Intelligence  42(1),  141--166 (2004)

\bibitem{Pinyol2013}
Pinyol, I., Sabater-Mir, J.: {Computational trust and reputation models for
  open multi-agent systems: A review}. Artificial Intelligence Review  40(1),
  1--25 (2013)

\bibitem{Ramchurn2004}
Ramchurn, S.D., Huynh, D., Jennings, N.R.: {Trust in multi-agent systems}. The
  Knowledge Engineering Review  19(01),  1--25 (2004)

\bibitem{Rasmusson1996}
Rasmusson, L., Jansson, S.: {Simulated social control for secure Internet
  commerce}. Proceedings of the 1996 workshop on New security paradigms - NSPW
  '96 pp. 18--25 (1996)

\bibitem{Sabater2005}
Sabater, J., Sierra, C.: {Review on computational trust and reputation models}.
  Artificial Intelligence Review  24(1),  33--60 (2005)

\bibitem{Ygge1997}
Sandholm, T., Ygge, F.: {On the Gains and Losses of Speculation in Equilibrium
  Markets}. In: Proceedings IJCAI'97. pp. 632--638. AAAI Press (1997)

\bibitem{Szabo1997}
Szabo, N.: {Formalizing and Securing Relationships on Public Networks.} (1997),
  \url{http://ojphi.org/ojs/index.php/fm/article/view/548/469}

\bibitem{Teutsch2017}
Teutsch, J., Reitwie{\ss}ner, C.: {A scalable verification solution for
  blockchains} (2017)

\bibitem{Vukolic2016}
Vukoli{\'{c}}, M.: {Hyperledger fabric: towards scalable blockchain for
  business}. Tech. Rep. Trust in Digital Life 2016, IBM Research (2016),
  \url{https://www.zurich.ibm.com/dccl/papers/cachin{\_}dccl.pdf}

\bibitem{Wood2014}
Wood, G.: {Ethereum: a secure decentralised generalised transaction ledger}.
  Ethereum Project Yellow Paper pp. 1--32 (2014)

\bibitem{Wooldridge2009}
Wooldridge, M.: {An Introduction to MultiAgent Systems}. Wiley Publishing, 2nd
  edn. (2009)

\bibitem{Zhou2008}
Zhou, R., Hwang, K., Cai, M.: {GossipTrust for Fast Reputation Aggregation in
  Peer-to-Peer Networks}. IEEE Transactions on Knowledge and Data Engineering
  20(9),  1282--1295 (2008)

\bibitem{Zyskind2016}
Zyskind, G.: {Efficient Secure Computation Enabled by Blockchain Technology}.
  Master thesis, Massachusetts Insitute of Technology (2016)

\end{thebibliography}
\bibliographystyle{splncs03}

\end{document}